\documentclass[12pt,a4paper]{article}
\usepackage[latin1]{inputenc}
\usepackage[T1]{fontenc}
\usepackage{amsmath}
\usepackage{amsfonts}
\usepackage{amssymb}
\usepackage[english]{babel}
\usepackage[dvips]{graphicx,color}
\title{A Dual Path Integral Representation for Finite Temperature
Quantum Field Theory}
\author{C.~Ccapa Ttira$^{a}$, C.~D.~Fosco$^{a}$,\\
A.~P.~C.~Malbouisson$^{b}$\\ and I.~Roditi$^{b}$\\ {\normalsize\it
$^a$Centro At\'omico Bariloche and Instituto Balseiro}\\
{\normalsize\it Comisi\'on Nacional de Energ\'\i a At\'omica} \\
{\normalsize\it 8400 Bariloche, Argentina.}\\ {\normalsize\it
$^b$Centro Brasileiro de Pesquisas F\'isicas - CBPF/MCT}\\
{\normalsize\it Rua Dr. Xavier Sigaud, 150, 22290-180 Rio de Janeiro,
RJ, Brazil}}
\begin{document}
\newcommand*{\be}{\begin{equation}} 
\newcommand*{\ee}{\end{equation}}
\newcommand*{\beq}{\begin{eqnarray}}
\newcommand*{\eeq}{\end{eqnarray}}
\def\d{{\rm d}}
\maketitle
\begin{abstract}
  \noindent We impose the periodicity conditions corresponding to the
  Matsubara formalism for Thermal Field Theory as constraints in the
  imaginary time path integral. These constraints are introduced by means
  of time-independent auxiliary fields which, by integration of the
  original variables, become dynamical fields in the resulting `dual'
  representation for the theory. This alternative representation has the
  appealing property of involving fields which live in one dimension less
  than the original ones, with a  quantum partition function whose
  integration measure is identical to the one of its classical counterpart,
  albeit with a different (spatially nonlocal) action. 
\end{abstract}
\section{Introduction}\label{sec:intro}
Quantum Field Theory (QFT) models with constrained configuration spaces
naturally arise within the context of modern applications, particularly gauge
invariant systems~\cite{teitelboim}. 

More recently, a formulation involving constraints has been also applied to
deal with different kinds of problems, namely, the static and
dynamical Casimir effects~\cite{kardar,emig,fm}, and to derive the overlap
Dirac operator~\cite{Neuberger1,Neuberger2} in a simpler way~\cite{Fosco}.
There, fields used to impose the constraints lead, after integrating out the
original variables, to an effective model where the dynamical fields live on the
constrained surface. Since the codimension of the boundaries is usually equal
to $1$, the effective model is defined in one dimension less than the
original one.

Here, we present the extension of that kind of approach to a less natural 
realm, that of QFT at finite temperature ($T>0$), to deal with the periodicity
constraints in the imaginary time. The foundations of QFT at $T>0$ were laid
down quite a long time ago~\cite{mats1,ume1}.  The original approach to this
topic, the
now called Matsubara (or `imaginary-time') formalism has been very
successful indeed in allowing for the evaluation of thermal effects in QFT,
both in High Energy~\cite{kap1} and Condensed Matter Physics.  

It allowed, for example, to study the new phenomena that emerge when using
a Statistical Mechanics description for quantum relativistic systems. It
also provided a convenient way to naturally extend the notion of Abelian
and non Abelian gauge fields, studying its consequences for particle
physics~\cite{kap1,le1,belac1}, in the $T > 0$ context.

A fundamental property introduced by this formalism is the imaginary-time
periodicity (antiperiodicity) conditions for the bosonic (fermionic) field
configurations in the path integral. That may be clearly seen already at
the level of the partition function, ${\mathcal Z}(\beta)$, for a system at
a temperature $T = 1/\beta$, with a Hamiltonian ${\widehat H}$:
\begin{equation}
{\mathcal Z}(\beta) \;=\; {\rm Tr} \big( e^{-\beta {\hat H}} \big)\,.
\end{equation}
Assuming first, for the sake of simplicity, that there is only one
(bosonic) degree of freedom, described by a coordinate $q$, the expression
above may be written more explicitly as follows:
\begin{equation}
{\mathcal Z}(\beta) \;=\; \int dq \; \langle q| e^{-\beta
H}|q\rangle\;=\; \int dq \; \langle q,-i\beta | q , 0 \rangle \;,
\end{equation}
where $|q,t\rangle$ denotes the usual `rotating basis' elements, which here
appear evaluated at imaginary values of $t$.  Then the standard path integral
construction for the transition amplitude between different times may be
applied, to obtain the partition function in the Matsubara formalism:
\begin{equation}\label{eq:defzh}
{\mathcal Z}(\beta) \;=\;
  \int_{\rm q(0)=q(\beta)}\, {\mathcal D} p \, {\mathcal D} q \;
  e^{\int_0^\beta d\tau \big[ i p \dot{q}
    \,-\, H(p,q)\big]} \;,
\end{equation}
where the measure includes phase-space paths $q(\tau), \, p(\tau)$ ($\tau \in
[0,\beta]$) such that $q(0) = q(\beta)$, while the $p(\tau)$ paths have free 
boundary conditions~\footnote{Note, however, that a slightly more symmetric form for
those conditions in the path integral for ${\mathcal Z}(\beta)$
could be used~\cite{teitelboim}.}.

When applied to a bosonic field theory in $d+1$ spacetime dimensions, this
procedure leads to field paths which are periodic in the imaginary time,
while the canonical momentum ones are, again, unrestricted. Moreover, when
the Hamiltonian is quadratic in the canonical momentum, integration of this
variable yields a model where the dynamical field is defined on $S^1 \times
R^{d}$, where the radius of $S^{1}$ is proportional to the inverse
temperature, $\beta$. In Fourier space, the corresponding frequencies
become the usual discrete Matsubara frequencies.

A characteristic feature of the Matsubara formalism (shared with the
real-time formulation) is that the introduction of a time dependence for
the fields seems to be unavoidable, even if one limits oneself to the
calculation of time independent objects.

With the aim of constructing a new representation where only static fields
are involved, we shall introduce here an alternative way of dealing with
$T>0$ QFT calculations. The procedure is inspired by a recent paper in
which a constrained functional integral approach is used to implement the
effect of fluctuating boundaries in the Casimir effect~\cite{kardar}.  In
the present context, this allows one to introduce the periodicity
conditions by means of Lagrange multipliers ($d$-dimensional when the field
lives in $d+1$ dimensions).  Then the original fields can be integrated,
what leaves a functional depending only on the $d$-dimensional Lagrange
multipliers.

This paper is organized as follows: in section~\ref{sec:method} we
introduce the method, using the harmonic oscillator as a convenient
framework. In section~\ref{sec:scalar} we deal with the real scalar field,
and in~\ref{sec:dirac} a Dirac field is considered. 
In section~\ref{sec:concl} we present our conclusions.
\section{The method}\label{sec:method}
\subsection{The periodicity constraint}
Let us see, again within the context of a system with a single degree of freedom,
how the thermal partition function may be obtained by imposing appropriate
constraints to the path integral for ${\mathcal Z}_0$, the (zero temperature) 
vacuum persistence amplitude.
For reasons that will become clear below, we start from its phase-space
path integral:
\begin{equation}\label{eq:defz0}
{\mathcal Z}_0 \;=\; \int \, {\mathcal D}p
  \,{\mathcal D}q \;
  e^{-{\mathcal S}[q(\tau),p(\tau)]} \;,
\end{equation}
where ${\mathcal S}$ is the first-order action, \mbox{${\mathcal S}
=\int_{-\infty}^{+\infty} d\tau \,{\mathcal L}$}, 
with ${\mathcal L} = - i p \dot{q} + H(p,q)$, 
and $H$ denotes the Hamiltonian, assumed to be of the form:
$H(p,q)=T(p)+V(q)$.

Of course, ${\mathcal Z}_0$ is the limit of an imaginary-time
transition amplitude,
\begin{eqnarray}
  {\mathcal Z}_0 &=& \lim_{T \to +\infty} \,
  \langle q_0, -i T | q_0, i T \rangle  \nonumber\\
  &=& \lim_{T \to +\infty} \,\sum_n | \langle q_0| n \rangle |^2 e^{-2 T
E_n}
  \; = \;\lim_{T \to +\infty} \, | \langle q_0| 0 \rangle |^2 e^{-2 T E_0}
\end{eqnarray}
where we have introduced $|n\rangle$, the eigenstates of ${\hat H}$, ${\hat
H}|n\rangle = E_n |n\rangle$, and $q_0$, the asymptotic value for $q_0$ at
$T \to \pm \infty$ (usually, $q_0 \equiv 0$). $E_0$ is the energy of
$|0\rangle$, the ground state.

Let us now see how one can write an alternative expression for ${\mathcal
Z}(\beta)$, by starting from the vacuum transition amplitude, ${\mathcal
Z}_0$, and imposing the appropriate constraints on the paths.  To that end,
we first use the superposition principle, introducing decompositions of the
identity at the imaginary times corresponding to $\tau = 0$ and $\tau = \beta$, 
so that we may write ${\mathcal Z}_0$ in the equivalent way:
\begin{equation}
{\mathcal Z}_0 \;=\; \lim_{T\to \infty}
\,\int dq_2 dq_1 \,
  \langle q_0, -i T | q_2, -i \beta \rangle \, \langle q_2, -i \beta | q_1 , 0
\rangle
  \, \langle q_1, 0 | q_0, i T \rangle \;,
\end{equation}
or, in a path integral representation,
\begin{eqnarray}\label{eq:aux1}
{\mathcal Z}_0 &=& \lim_{T \to \infty} \;
  \int dq_2 dq_1 \, \int_{q(\beta)= q_2}^{q(T) = q_0} \,
  {\mathcal D}p \,{\mathcal D}q \; e^{- \int_\beta^T d\tau {\mathcal L}}
  \nonumber\\
  &\times& \int_{q(0)=q_1}^{q(\beta)=q_2} \, {\mathcal D}p \,{\mathcal
D}q\;
  e^{- \int_0^\beta d\tau {\mathcal L}}
  \;\int_{q(-T)=q_0}^{q(0)=q_1} \, {\mathcal D}p \,{\mathcal D}q\;
  e^{- \int_{-T}^0 d\tau {\mathcal L}} \;.
\end{eqnarray}
The representation above is quite useful in order to understand which is
the correct way to impose the constraints, to obtain ${\mathcal
Z}(\beta)$.  
In short, to reproduce ${\mathcal Z}(\beta)$ we have to impose periodicity
constraints for {\em both\/} phase space variables. Indeed, let us
introduce an object ${\mathcal Z}_s(\beta)$ that results from imposing those
constraints on the ${\mathcal Z}_0$ path integral, and extracting a 
${\mathcal Z}_0$ factor:
\begin{equation}
{\mathcal Z}_s(\beta) \;\equiv \; \frac{\int \,{\mathcal
D}p \,{\mathcal D}q \;\delta\big( q(\beta) - q(0)\big)\,
\delta\big(p(\beta) -
    p(0)\big) \, e^{-{\mathcal S}}}{\int \,{\mathcal D}p\,{\mathcal D}q \,
e^{-{\mathcal S}}}
  \,.
\end{equation}
Then, the use of the superposition principle yields:
$$
\int \, {\mathcal D}p \,{\mathcal D}q \; \delta\big( q(\beta) - q(0)\big)
\delta\big( p(\beta) - p(0)\big) \; e^{-{\mathcal S}}= \lim_{T\to \infty}
\,\int
dp_1
dq_1 \,\Big[
$$
\begin{equation}
  \langle q_0, -i T | p_1, -i \beta  \rangle \,
  \langle p_1, -i \beta | q_1, -i \beta \rangle \,
  \langle q_1 , -i \beta | q_1 , 0 \rangle \,
  \langle q_1 , 0 | p_1 , 0 \rangle\,
  \langle p_1 , 0 | q_0 , i T \rangle \Big]
\end{equation}
or
$$
\int \, {\mathcal D}p \,{\mathcal D}q \; \delta\big( q(\beta) - q(0)\big)
\delta\big( p(\beta) - p(0)\big) e^{-{\mathcal S}} \,=\, \lim_{T\to \infty}
\,e^{ -
E_0
  ( 2 T - \beta)}
$$
$$
\times \, \int \frac{dp_1 dq_1}{2\pi} \, \langle q_0|0\rangle \langle 0 |
p_1\rangle \, \langle q_1 , -i \beta | q_1 , 0 \rangle \, \langle p_1|0\rangle
\langle 0 | q_0\rangle
$$
$$
  = \, \lim_{T\to \infty} \,e^{ - E_0 ( 2 T - \beta)}
  \,|\langle q_0 | 0 \rangle |^2 \,\int dq_1 \, \langle q_1 , -i \beta | q_1 ,
0
  \rangle .
$$
\begin{equation}
	= {\mathcal Z}_0 \, \times \, e^{\beta E_0} \,{\mathcal Z}(\beta) 
	\,=\, {\mathcal Z}_0 \, \times \, {\rm Tr} \big[ e^{-\beta ({\hat
	H} - E_0) } \big] \,.
\end{equation}	
Then we conclude that
\begin{equation}
{\mathcal Z}_s(\beta) \;=\; {\rm Tr} \big[ e^{-\beta \,
    :{\hat H}:} \big]
\end{equation}
where $:{\hat H}:$ denotes the normal-ordered Hamiltonian operator, i.e.:
\begin{equation}
  :{\hat H}: \;\equiv\;   {\hat H} \, - \, E_0  \,.
\end{equation}
The conclusion is that, by imposing periodicity on both phase space
variables, and discarding $\beta$-independent factors (since they would be
canceled by the normalization constant) we obtain ${\mathcal Z}_s(\beta)$,
the partition function
corresponding to the original Hamiltonian, the ground state energy redefined
 to zero. The subtraction of the vacuum energy is
usually irrelevant (except in some exceptional situations), as it is wiped
out when taking derivatives of the free energy to calculate physical
quantities.

Note that the introduction of periodicity constraints for both variables
is not in contradiction with the usual representation, (\ref{eq:defzh}),
where they only apply to $q$, since they corresponds to different sets of
paths. Indeed, in our approach the new constraints are crucial in order to
get rid of the unwelcome factors coming from paths which are outside
of the $[0,\beta]$ interval (which are absent
from the standard path integral).

We conclude this derivation of the boundary conditions by showing explicitly
why the usual procedure of introducing a periodicity constraint for just the
coordinate $q(\tau)$ would not be sufficient. Indeed, we can see that
$$
\int \, {\mathcal D}p \,{\mathcal D}q \; \delta\big( q(\beta) - q(0)\big)
\exp
\big\{-{\mathcal S}[q(\tau),p(\tau)] \big\}
$$
\begin{equation}
  \; = \; \lim_{T\to \infty} \,e^{ - E_0 ( 2 T - \beta)} \,
  \int dq_1 |\langle 0 | q_1 \rangle |^2 \,
  \langle q_1 , -i \beta | q_1 , 0 \rangle \;,
\end{equation}
and taking the ratio with the (unconstrained) vacuum functional,
\begin{equation}
 \frac{\int \,{\mathcal D}p \,{\mathcal D}q \;\delta\big(
q(\beta) - q(0)\big)\, e^{-{\mathcal S}}}{\int \,{\mathcal D}p\,{\mathcal
D}q \,
e^{-{\mathcal S}}} \, = \, e^{ \beta E_0 } \, \int dq_1 |\langle 0 | q_1
\rangle |^2
\, \langle q_1 , -i \beta | q_1 , 0 \rangle \;,
\end{equation}
where we cannot extract a ${\mathcal Z}(\beta)$ factor, due to the presence
of the squared vacuum wave function inside the integral.
It is not difficult to realize that that factor, whose entanglement makes
it impossible to extract the partition function, is due to
contributions from paths outside of the $(0,\beta]$ interval.

Summarizing, we have shown that the proper way to extract the partition
function {\em from the $T=0$ partition function\/} ${\mathcal Z}_0$, is to
impose periodicity constraints for both the coordinate and its canonical
momentum, a procedure that yields a ${\mathcal Z}_0$ factor times the
thermal partition function, ${\mathcal Z}_s(\beta)$. We present, in Figure
1, a pictorial representation of this `compactification' mechanism. 
\begin{figure}
 \centering
 \includegraphics[angle=-90,scale=0.65]{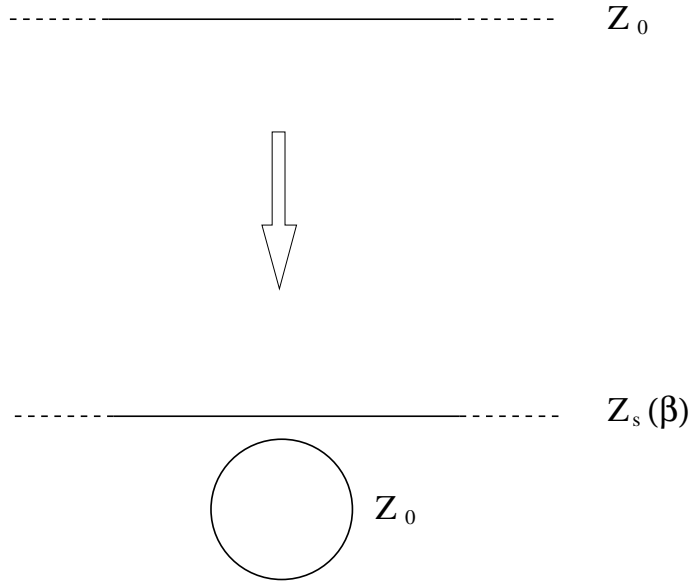}
 \caption{Representation of the compactification mechanism}
 \label{fig:fig1}
\end{figure}

\subsection{Auxiliary fields}
Let us now see how the use of auxiliary fields to exponentiate the constraints
leads naturally to an alternative representation.  The two $\delta$-functions
require the introduction of two auxiliary fields, $\xi_1$ and $\xi_2$.
which are just real (time-independent) variables in this case. Using
the notation $Q_1 \equiv q$ and $Q_2 \equiv p$, we have 
\begin{equation}
	\prod_{a=1}^2\Big\{ \delta\big[ Q_a(\beta) - Q_a(0) \big] \Big\} 
\;=\; \int \frac{d^2\xi}{(2\pi)^2} \, e^{ i \sum_{a=1}^2 \xi_a 
\big[ Q_a(\beta) - Q_a(0) \big]} \;.
\end{equation}
Using this representation for the constraints and interchanging the order of 
integration for the multipliers and the phase space variables, the resulting
expression for ${\mathcal Z}_s(\beta)$ may be written as follows:
\begin{eqnarray}\label{eq:zbd}
{\mathcal Z}_s(\beta) &=& {\mathcal N}^{-1}
  \,\int_{-\infty}^{\infty}
  \frac{d\xi_1}{2\pi} \int_{-\infty}^{\infty}\frac{d\xi_2}{2\pi}
  \int {\mathcal D} Q
  \nonumber\\
  &\times& e^{- {\mathcal S}(Q) \,+\,i\, \int_{-\infty}^\infty
d\tau  j_a(\tau) Q_a(\tau) } \;,
\end{eqnarray}
where $\mathcal N \equiv {\mathcal Z}_0$, and we have introduced the
notation:
\begin{equation}\label{eq:defja}
j_a(\tau) \equiv \xi_a \,\big[ \delta(\tau-\beta) - \delta(\tau) \big] \;. 
\end{equation}
The phase-space measure has been written in terms of $Q$:
\begin{equation}
 {\mathcal D} Q \equiv \prod_{-\infty < \tau < \infty} \frac{dq(\tau)
dp(\tau)}{2 \pi}  \;.
\end{equation}
For the particular case of a harmonic oscillator with unit mass and 
frequency $\omega$, we have 
\begin{equation}\label{eq:zb2}
{\mathcal S} (Q) \;=\; {\mathcal S}_0(Q)\,=\, \frac{1}{2} 
\int_{-\infty}^{+\infty} d\tau \, Q_a(\tau)
\widehat{{\mathcal K}}_{ab} Q_a(\tau) \;,
\end{equation}
where $\widehat{\mathcal K}_{ab}$ are the elements of the $2 \times 2$
operator matrix $\widehat{\mathcal K}$, given by:
\begin{equation}\label{eq:defk}
  \widehat{\mathcal K} \;=\;
  \left(
    \begin{array}{cc}
      \omega^2              & i \frac{d}{d\tau} \\
      - i \frac{d}{d\tau}     & 1
    \end{array}
  \right) \;.
\end{equation}
Thus the integral over $Q$ is a Gaussian; it may therefore be written as
follows:
\begin{equation}
{\mathcal Z}_s(\beta) \;=\; 2 \pi \, {\mathcal N}^{-1}\, 
\big(\det \widehat{\mathcal K}\big)^{-\frac{1}{2}} \; \int
  \frac{d^2\xi}{(2\pi)^2} \,\, e^{- \frac{1}{2} \xi_a M_{ab} \xi_b } \;,
\end{equation}
with
\begin{equation}\label{eq:defm}
  M \;\equiv\; \Omega(0_+)\,+\Omega(0_-) \,-\, \Omega(\beta) \,-\, \Omega(-\beta) \;,
\end{equation}
where $\Omega(\tau)$ denotes the inverse of the operator ${\mathcal K}$ of
(\ref{eq:defk}); namely,
\begin{equation}
 \widehat{\mathcal K}_{ac}\Omega_{cb}(\tau) \;=\; \delta_{ab} \, \delta(\tau) 
\end{equation}
where the $\Omega_{ab}$'s denote the matrix elements of $\Omega$.

The explicit form of this object may be easily found to be the following:
\begin{equation}\label{eq:defom}
  \Omega(\tau)\;\equiv\;
  \left(
    \begin{array}{cc}
      \frac{1}{2\omega} & \frac{i}{2} {\rm sgn}(\tau) \\
      -\frac{i}{2} {\rm sgn}(\tau) & \frac{\omega}{2}
    \end{array}
  \right) \; e^{-\omega |\tau|}
\end{equation}
(${\rm sgn} \equiv$ sign function).

Equation (\ref{eq:defom}) can be used in (\ref{eq:defm}), to see that:
\begin{eqnarray}\label{eq:m1}
	M &=&  \big[\Omega(0_+)\,+\Omega(0_-) \big] \, ( 1 -
	e^{-\beta\omega} ) \nonumber\\
&=&
\left(
 \begin{array}{cc}
 \omega^{-1} & 0 \\
 0 & \omega
    \end{array}
  \right) \; (n_B(\omega) + 1)^{-1} \;,
\end{eqnarray}
where
\begin{equation}
  n_B(\omega) \equiv (e^{\beta \omega} - 1)^{-1}\
\end{equation}
is the Bose-Einstein distribution function (with the zero of energy set at the ground state).

Finally, note that ${\mathcal N}$ exactly cancels the  
$\big(\det \widehat{\mathcal K}\big)^{-\frac{1}{2}}$ factor, and thus
we arrive to a sort of `dual' description for the partition
function, as an integral over the $\xi_a$ variables:
\begin{equation}\label{eq:zosc}
{\mathcal Z}_s(\beta) \;=\;
  \int \frac{d^2\xi}{2\pi} \,
  e^{- \frac{\omega^{-1} \, \xi_1^2 \,+\;
      \omega \; \xi_2^2}{2 [ n_B(\omega) + 1 ]}} \;.
\end{equation}
This integral is over two real variables $\xi_a$, which are $0$-dimensional
fields, one dimension less than the $0+1$ dimensional original theory. To
interpret this integral we may compare it with the one corresponding to the 
classical statistical mechanics version of this system. 
To that end, we evaluate the partition function in the classical
(high-temperature) limit.  In that limit, we approximate the integrand accordingly 
to see that
${\mathcal Z}_s(\beta)$ becomes:
\begin{equation}\label{eq:zosc2}
{\mathcal Z}_s(\beta) \;\simeq\;\int
  \frac{d^2\xi}{2\pi} \,
  e^{- \beta H(\xi_1,\xi_2)} \;\;\; (\beta << 1) \;,
\end{equation}
where:
\begin{equation}
  H(\xi_1,\xi_2) \;\equiv\; \frac{1}{2} \big( \xi_1^2 \,+\, \omega^2
\xi_2^2
  \big)\;.
\end{equation}
We see that (\ref{eq:zosc2}) corresponds exactly to the classical partition
function for a harmonic oscillator, when the identifications: $\xi_1=p$
(classical momentum), and $\xi_2=q$ (classical coordinate) are made
\begin{equation}\label{eq:zosc3} {\mathcal Z}_s(\beta) \;\simeq\;\int
\frac{dp
    dq}{2\pi} \,
  e^{- \beta \frac{1}{2} \big( p^2 \,+\, \omega^2 q^2\big)}
  \;\;\; (\beta << 1) \;.
\end{equation}

On the other hand, had the exact form of the integral been kept (no
approximation), we could still have written an expression similar to the
classical partition function, albeit with an `effective Hamiltonian'
$H_{eff}(\xi_1,\xi_2)$:
\begin{equation}\label{eq:zosc4}
{\mathcal Z}_s(\beta) \;=\;\int
  \frac{d^2\xi}{2\pi} \,
  e^{- \beta H_{eff}(\xi_1,\xi_2)} \;,
\end{equation}
where:
\begin{equation}
  H_{eff}(\xi_1,\xi_2) \;\equiv\; \frac{1}{2 \beta} \,
  \big( n_B(\omega) + 1 \big)^{-1}\, \big(\omega^{-1} \, \xi_1^2 \,+\;
  \omega \; \xi_2^2 \big) \;.
\end{equation}
This shows that the quantum partition function may also be written as a
classical one, by using a $\beta$-dependent Hamiltonian, which of course
tends to its classical counterpart in the high-temperature limit.  

By integrating out the auxiliary fields in the (exact) expression for
the partition function (\ref{eq:zosc}), we obtain:
\begin{equation}\label{eq:free}
{\mathcal Z}_s(\beta) \;=\; n_B(\omega)
\,+\,
  1 \;=\;
  \frac{1}{1\,-\, e^{-\beta \omega}} \;.
\end{equation}
which is the correct result. In what follows, to simplify the notation, we
shall omit writing the `$s$' subscript in ${\mathcal Z}(\beta)$, assuming
implicitly that one is dealing with the normal-ordered Hamiltonian.

An important fact that has emerged from an analysis of the classical
(high-temperature) limit: the auxiliary fields do have a physical
interpretation.  The multiplier associated to the periodicity condition for
$q$ plays the role of a classical momentum, while the one corresponding to
the periodicity for the momentum becomes a generalization of the classical
coordinate. The same interpretation might also be retained far from the
classical limit, but then the Hamiltonian departs from the classical one,
receiving quantum corrections.

This representation is also valid for interacting theories. To that
effect, note that, even when the action ${\mathcal S}$ is not quadratic, we
may still give a formal expression for the alternative representation.
Indeed, denoting by 
${\mathcal Z}(J)$ the {\em zero-temperature\/} generating functional of
correlation functions of the canonical variables:
\begin{equation}
{\mathcal Z}(J) \;=\; \int {\mathcal D}Q \, e^{ - {\mathcal S}(Q) + 
\int_{-\infty}^{\infty}d\tau J_a(\tau) Q_a(\tau)}
\end{equation}
and by ${\mathcal W}(J)$ the corresponding functional for connected ones, 
we see that
\begin{equation}
{\mathcal Z}_s(\beta)\,=\, [{\mathcal Z}(0)]^{-1} \int \frac{d^2 \xi}{(2\pi)^2} 
\, \exp \{{\mathcal W}\big[i \, j(\tau)\big]\} \;,
\end{equation}
where, with our normalization conventions, ${\mathcal Z}(0) = {\mathcal
Z}_0$ (the vacuum functional for the interacting case).

Thus, a possible way to derive the effective Hamiltonian in the interacting
case is to obtain first ${\mathcal W}[J]$, and then to replace the
(arbitrary) source $J(\tau)$ by $i j(\tau)$, where $j(\tau)$ is the function 
of the auxiliary field defined in (\ref{eq:defja}).
Of course, ${\mathcal W}$ cannot be obtained exactly, except in
very special cases. Otherwise, a suitable perturbative expansion can be
used. In any case, ${\mathcal W}$ can be functionally expanded in powers
of the source $J(\tau)$:
\begin{equation}
{\mathcal W}[J] \,=\, {\mathcal W}[0] \,+\, \sum_{n=1}^\infty \frac{1}{n!}
\int_{\tau_1,\ldots,\tau_n} {\mathcal W}_{ab}^{(n)}(\tau_1,\ldots, \tau_n)
J_{a_1}(\tau_1) \ldots  J_{a_n}(\tau_n)
\end{equation}
where each coefficient ${\mathcal W}^{(n)}$ is the $n$-point connected
correlation function. The expansion above immediately yields an expansion
for $H_{eff}$ in powers of the auxiliary fields. It is important to stress
that this expansion is not necessarily a perturbative expansion. 
Indeed, the strength of each term is controlled by ${\mathcal W}^{(n)}$,
which could even be exact (non-perturbative) in a coupling constant. To fix
ideas, let us see what happens when one keeps only up to the $n=4$ term,
assuming also that there is ${\mathcal Q}_a \to - {\mathcal Q}_a$
symmetry in ${\mathcal S}$. Then, we first see that the ${\mathcal W}[0]$ 
is cancelled by the ${\mathcal N}$ factor, and on the other hand we obtain
\begin{equation}
{\mathcal Z}_s(\beta) \;=\;\int
  \frac{d^2\xi}{(2\pi)^2} \,
  e^{- \beta H_{eff}(\xi_1,\xi_2)} \;,
\end{equation}
where
\begin{eqnarray}
H_{eff} &=& \frac{1}{2\beta}
\int_{\tau_1,\tau_2} {\mathcal W}_{a_1a_2}^{(2)}(\tau_1,\tau_2)
j_{a_1}(\tau_1) j_{a_2}(\tau_2) \nonumber\\
& - & \frac{1}{4! \beta}
\int_{\tau_1,\tau_2} {\mathcal
W}_{a_1 a_2 a_3 a_4}^{(2)}(\tau_1,\tau_2,\tau_3,\tau_4)
j_{a_1}(\tau_1) j_{a_2}(\tau_2) j_{a_3}(\tau_3) j_{a_4}(\tau_4)\nonumber\\
&+& \ldots 
\end{eqnarray}
Using the explicit form of $j_a(\tau)$ in terms of the auxiliary fields, we
see that:
\begin{equation}
 H_{eff} \,=\,  H_{eff}^{(2)} \,+\,  H_{eff}^{(4)} \,+\, \ldots   
\end{equation}
where
\begin{eqnarray}
H_{eff}^{(2)} &=& \frac{1}{2}  {\mathcal M}^{(2)}_{ab} \xi_a \,
\xi_b \nonumber\\
H_{eff}^{(4)} &=&  \frac{1}{4!} {\mathcal M}^{(4)}_{abcd}\; \xi_{a}\,
\xi_{b}\, \xi_{c} \, \xi_{b}\nonumber\\
\ldots &=& \ldots \nonumber\\
H_{eff}^{(2k)} &=&  \frac{1}{(2 k)!} {\mathcal M}^{(2k)}_{a_1
\ldots a_{2k}}\; \xi_{a_1}\ldots \xi_{a_{2k}}\;,
\end{eqnarray}
where the explicit forms of the coefficients ${\mathcal M}^{(2k)}$ in terms
of ${\mathcal W}^{(2k)}$ may be found, after some algebra. For example ${\mathcal M}^{(2)}$ is a
diagonal matrix:
\begin{equation}
 {\mathcal M}^{(2)}\,=\,
\left( \begin{array}{cc}
c_1 & 0 \\
0   & c_2  
\end{array}
\right)
\end{equation}
where
\begin{equation}
c_a = \frac{1}{\beta}\, \int \frac{d\nu}{\pi} \;
\big( 1 - e^{ - i \nu \beta} \big)\, \tilde{\mathcal W}_{aa}(\nu) 
\end{equation}
(where the tilde denotes Fourier transform). It is immediate to realize that 
$c_1$ plays the role of an effective coefficient for the
kinetic term ($\propto p^2$) in the effective Hamiltonian, while $c_2$ does
introduce an effective quadratic potential. Note that they will, in general,
depend on $\beta$, $\omega$, and on any additional coupling constant the
system may have. For the harmonic oscillator case we have the rather simple
form:
\begin{eqnarray}
 c_1 &=&  \frac{1}{\omega ( n_B(\omega) + 1)} \nonumber\\
 c_2 &=&  \frac{\omega}{ n_B(\omega) + 1} \;.
\end{eqnarray}
The quartic term involves ${\mathcal M}^{(4)}$, which may be written in terms
of the connected $4$-point function:
$$
 {\mathcal M}^{(4)}_{a b c d} \;=\; \frac{1}{\beta}  \,
\Big[ - {\mathcal W}^{(4)}_{a b c d}(0,0,0,0) \,+\, 
4 \,{\mathcal W}^{(4)}_{a b c d}(\beta,\beta,\beta,0) 
$$
\begin{equation}
\,-\,  6 \, {\mathcal W}^{(4)}_{a b c d}(\beta,\beta,0,0)  
\,+\, 4 \, {\mathcal W}^{(4)}_{a b c d}(\beta,0,0,0)
\,-\, {\mathcal W}^{(4)}_{a b c d}(0,0,0,0) \Big]_{sym}
\end{equation}
where the $sym$ suffix denotes symmetrization under simultaneous interchange
of time arguments and discrete indices. Of course, this expression could also
be written in Fourier space; we shall leave the analog of this term 
for the case of the real scalar field.

\subsection{Generating functional}
Now we proceed to the calculation of thermal correlation functions within
the approach that we developed for the calculation of the partition function.
To that end, we shall introduce the generating functional of correlation
functions, to be denoted by ${\mathcal Z}_s(\beta , J)$,
and ${\mathcal W}_s(\beta, J) \equiv \ln {\mathcal Z}_s(\beta,J)$  the
generating functional of connected correlation functions. With these
conventions, the connected correlation functions are given by:
\begin{equation}
	\langle Q_{a_1}(\tau_1) \ldots Q_{a_n}(\tau_n) \rangle_{conn} \;=\;
 \Big[\frac{\delta^n {\mathcal W}_s(\beta, J)}{\delta
J_{a_1}(\tau_1)\,\ldots\,
 \delta J_{a_n}(\tau_n)}\Big]_{J=0}\;.
\end{equation}

From the previous subsection, we know that path integral expression for the
generating functional shall be:
\begin{equation}
{\mathcal Z}_s(\beta , J) \,=\, {\mathcal N}^{-1}
  \; \int \frac{d^2\xi}{2\pi} \; \int {\mathcal D}Q
\, e^{- {\mathcal S}(Q) \,+\,\int_{-\infty}^\infty
d\tau  Q_a(\tau) (J_a(\tau) + i j_a(\tau))}
\end{equation}
where ${\mathcal S}(Q)$ denotes the first-order form of the action.

Let us evaluate ${\mathcal Z}_s(\beta , J)$ explicitly for the case of the
harmonic oscillator, where ${\mathcal S}$ is a quadratic form.
The result of performing the Gaussian integral over $Q$, may be written in
this case as follows:
\begin{equation}\label{eq:genf2}
{\mathcal Z}_s(\beta , J)=
\int  \frac{d^2\xi}{2\pi} e^{\frac{1}{2} \int_{\infty}^{+\infty}
d\tau_1\int_{\infty}^{+\infty} d\tau_2 \, (J_a(\tau_1) + i j_a(\tau_1))
    \Omega_{ab}(\tau_1,\tau_2) (J_b(\tau_2) + i j_b(\tau_2))} \;.
\end{equation}
 
Using the explicit form for $j_a$, we see that the expression above is
equivalent to:
\begin{equation}
{\mathcal Z}_s(\beta , J) \; =\; 
{\mathcal Z}_s(J)
\int  \frac{d^2\xi}{2\pi} e^{-\frac{1}{2} \xi_a M_{ab} \xi_b 
\,+\, i \xi_a N_a}
\end{equation}
where ${\mathcal Z}_s(J)$ is the zero-temperature generating functional, 
\begin{equation}
{\mathcal Z}_s(J)=
e^{\frac{1}{2} \int_{\infty}^{+\infty}
d\tau_1\int_{\infty}^{+\infty} d\tau_2 \, J_a(\tau_1)
    \Omega_{ab}(\tau_1,\tau_2) J_b(\tau_2)}
\end{equation}
and
\begin{equation}\label{eq:defna}
 N_a \,\equiv \, \int_{-\infty}^{+\infty} d\tau 
\big[ 
(\hat{\mathcal K}^{-1})_{ab}(\beta,\tau) - (\hat{\mathcal
K}^{-1})_{ab}(0,\tau)\big] J_b(\tau) \;.
\end{equation}
Integrating out the auxiliary fields, and recalling the results of the
previous subsection,
\begin{equation}\label{eq:ecuaZJ}
Z_s(\beta,J)\, = \,  Z(\beta) \; Z_s(J) 
\; e^{-\frac{1}{2} N_a \, [M^{-1}]_{ab} N_b}
\end{equation}
where the $N_a$ are the functionals of $J_a$ defined in (\ref{eq:defna}).

Neglecting a source-independent term (irrelevant for the calculation of
correlation functions), the ${\mathcal W}$ generating
functional will have the structure,
\begin{equation}
{\mathcal W}_s(\beta,  J) \,=\, \frac{1}{2} \int d\tau_1 \int d\tau_2 \, 
J_a(\tau_1) G_{ab} (\tau_1, \tau_2) J_b(\tau_2) 
\end{equation}
where $G_{ab}$ denotes the thermal correlation function:
\begin{equation}
G_{ab}(\tau_1,\tau_2) =  G_{ab}(\tau_1-\tau_2)
= \langle Q_a(\tau_1) Q_b(\tau_2) \rangle \;,
\end{equation}
whose structure we shall now write more explicitly.  We first note that:
\begin{equation}
 G_{ab}(\tau_1, \tau_2) \;=\; G^{(0)}_{ab}(\tau_1, \tau_2)
-  U^{(\beta)}_{ab}(\tau_1,\tau_2) 
\end{equation}
where $G^{(0)}$ is the zero-temperature correlation function
while $U^{(\beta)}$ denotes a temperature-dependent
piece.

We can write a more explicit form for the two terms that enter into the
expression above for the correlation function. Indeed, for $G_0^{(\beta)}$ 
we have:
\begin{equation}
 G^{(0)}(\tau) \;=\; \frac{e^{- \omega |\tau_1 - \tau_2| }}{2 \omega} \, 
\left( 
\begin{array}{cc}
1 &  i \omega  \\
- i \omega  & \omega^2 
\end{array}
\right)
\end{equation}
while for $U^{(\beta)}$ we may use the following matrix representation:
\begin{equation}\label{eq:hbeta1}
	U^{(\beta)} \,=\, \big[{\mathcal K}^{-1}(\tau_1,\beta) - {\mathcal K}^{-1}(\tau_1,0) \big] 
	M^{-1} \big[{\mathcal K}^{-1}(\beta, \tau_2) -
	{\mathcal K}^{-1}(0,\tau_2) \big] 
\end{equation}	
where ${\mathcal K}^{-1}(\tau,\tau')$ denotes the kernel of the inverse of
$\hat{\mathcal K}$. It is clear that $G^{(0)}$ is invariant under
a translation in both time arguments. So is $U^{(\beta)}$, but one has to
carry on the calculation in (\ref{eq:hbeta1}) to see that explicitly;
indeed, assuming that both time arguments lie between $0$ and $\beta$, we
have:
\begin{equation}
U^{(\beta)}(\tau_1,\tau_2) \;=\;  - \frac{1}{2} \, n_B(\omega)
\; 
	\left( 
	\begin{array}{cc}
	\frac{1}{\omega} ( e^{-\omega\tau} +
	e^{\omega \tau} ) & i  ( 
	e^{-\omega\tau} - e^{\omega\tau}) \\
	-i  ( e^{-\omega\tau} -
	e^{\omega\tau} ) &  \omega (
	e^{-\omega\tau} + e^{\omega\tau})
	\end{array}
	\right)
\end{equation}	
where $\tau \equiv \tau_1 - \tau_2$. 

Thus, the thermal propagator only depends on $\tau$, the difference between
the time arguments, and it is naturally defined on $[-\beta,\beta]$.

One is usually interested in the $\langle q q \rangle$ correlation function,
which here we can immediately read off the general expressions above, since
it corresponds to the $11$ matrix element. For $0 < \tau \leq \beta$, we
may derive the simpler expression:
\begin{equation}\label{eq:ecuapropag1}
G_{11}(\tau ) =
\frac{1}{2w}[ ( 1 + n_B(w)) e^{-w \tau} + n_B(w)e^{w\tau} ]\;,
\end{equation}
which is the correct result~\footnote{See, for example,  expression (2.32)
on page 23 of \cite{belac1}.}. 
This is the unique solution to the differential equation:
\begin{equation}
( -\partial_\tau^2 + \omega^2 ) \, G_{11}(\tau ) \;=\; \delta (\tau) 
\end{equation}
subject to the condition $G_{11}(\tau-\beta) = G_{11}(\tau)$. Moreover, we can
understand that, in our construction, the zero-temperature part appears
naturally as the propagator for the unconstrained system. On the other
hand, $U^{(\beta)}$ is a solution to the homogeneous version of the equation
above, and emerges as due to the boundary conditions.

So far, we have seen that the previous form correctly reproduces the
free $2$-point function, for all the phase space variables, when both time
arguments are inside the $[0,\beta]$ interval. A somewhat lengthy
calculation shows that $G$ {\em vanishes\/} when one of its arguments is
outside of that interval while the other is inside, and that it coincides with
the $T=0$ function, $G^{(0)}$, when both are outside. Besides, in the last
case, the $[0,\beta]$ interval is bypassed. More explicitly:
\begin{equation}\label{eq:gcases}
	G(\tau_1,\tau_2) \;=\; 
	\left\{ 
	\begin{array}{ccccc}
		0 & {\rm if} & \tau_1 > \beta &{\rm and}& 0 < \tau_2 < \beta \\
		G^{(0)}(\tau_1,\tau_2) & {\rm if} & \tau_1 > \beta &{\rm
		and}& \tau_2 > \beta \\
		G^{(0)}(\tau_1,\tau_2) & {\rm if} & \tau_1 < \beta &{\rm
		and}& \tau_2 < \beta \\
		G^{(0)}(\tau_1-\beta,\tau_2) & {\rm if} & \tau_1 > \beta &{\rm
		and}& \tau_2 < 0 
	\end{array} \right. \;,
\end{equation}
and the remaining cases follow from  Bose symmetry. 

This is an important property, since it allows us to perform a consistency check: 
indeed, coming back to the interacting case we
considered in the previous subsection, one could have decided to perform
the integrals in the opposite order, namely, first over the auxiliary
fields and afterwards over $Q_a$. Then, the perturbative expansion would have
involved the evaluation of Gaussian averages of the integration term, via
Wick's
theorem, with $G$ as the fundamental contraction. Since the interaction
term involves an infinite time interval, one produces not only the
perturbative corrections to the partition function (when both times are
inside $[0,\beta]$), but also a contribution that takes care of the normal order 
of the Hamiltonian (namely, when both arguments are outside of $[0,\beta]$.

\section{Scalar field}\label{sec:scalar}
The extension of the harmonic oscillator results to the QFT of a real scalar
field $\varphi$ in $d+1$ (Euclidean)
dimensions is quite straightforward. Let $\varphi(x)=\varphi(\tau,{\mathbf
  x})$ where $x=(\tau,{\mathbf x}) \in {\mathbb R}^{(d+1)}$, $\tau \in
{\mathbb R}$ and ${\mathbf x} \in {\mathbb R}^{(d)}$. Proceeding along the
same lines as for the case of a single degree of freedom, we consider first
the free case.
\subsection{Free partition function}\label{ssec:freepart}
The free Euclidean action in terms of the phase-space variables ${\mathcal
S}_0$, is in this case given by:
\begin{equation}\label{eq:defs0}
  {\mathcal S}_0 \,=\,\int d^{d+1}x \,\Big[ - i \pi \partial_\tau \varphi 
  + {\mathcal H}_0(\pi,\varphi) \Big] \;,
\end{equation}
with
\begin{equation} {\mathcal H}_0(\pi,\varphi) \;\equiv\; \frac{1}{2} \Big[
  \pi^2 \,+\, |{\mathbf \nabla}\varphi |^2 + m^2 \varphi^2 \Big] \;.
\end{equation}

We then have to implement the periodic boundary conditions both for
$\varphi(\tau,{\mathbf x})$ and its canonical momentum $\pi(\tau,{\mathbf
x})$
\begin{equation}
  \varphi\left(\beta,{\mathbf x}\right) \,=\, \varphi \left(0,{\mathbf
      x}\right)\;,\;\;\; \pi\left(\beta,{\mathbf x}\right) \,=\,
\pi\left(0,{\mathbf
      x}\right)\;,\;\;\forall\, {\mathbf x} \, \in {\mathbb R}^{(d)} \;,
\end{equation}
which requires the introduction of two {\em time-independent\/} Lagrange
multiplier fields: $\xi_a({\mathbf x})$, $a=1,\,2$.  Defining a
two-component field \mbox{$\Phi = (\Phi_a)$}, \mbox{$a=1,\,2$}, such that
$\Phi_1 = \varphi$ and $\Phi_2 = \pi$, an analogous procedure to the one
followed for the harmonic oscillator yields, for the free partition
function ${\mathcal Z}_0(\beta)$:
\begin{equation}\label{eq:zesc1} 
{\mathcal Z}_0(\beta) \;=\; {\mathcal N}^{-1}
  \,\int \,
  {\mathcal D}\xi \, \int {\mathcal D}\Phi \; 
  e^{-\frac{1}{2} \int d^{d+1}x \, \Phi_a 
    {\hat{\mathcal K}}_{ab} \Phi_b \,+\,i\, \int d^{d+1}x  j_a \Phi_a}
  \;,
\end{equation}
where $j_a(x) \equiv \xi_a({\mathbf x}) \big[ \delta(\tau-\beta) -
\delta(\tau) \big]$ and:
\begin{equation}\label{eq:newdefk} 
{\widehat{\mathcal K}}\;=\;
  \left( 
    \begin{array}{cc}
      {\hat h}^2       & i \frac{\partial}{\partial\tau} \\
      - i \frac{\partial}{\partial\tau}     & 1 
    \end{array}
  \right) \;,
\end{equation}
where we have introduced \mbox{${\hat h} \equiv \sqrt{-\nabla^2 + m^2}$},
the first-quantized energy operator for massive scalar particles.
Performing the integral over $\Phi$, yields the partition function in terms
of the Lagrange multipliers:
\begin{equation} 
{\mathcal Z}_0(\beta) \;=\; \int {\mathcal D}\xi \,e^{-
    \frac{1}{2} \int d^dx \int d^dy \, \xi_a ({\mathbf x}) \; \langle
{\mathbf
      x} | {\hat M}_{ab}| {\mathbf y} \rangle \; \xi_b ({\mathbf y})} \;,
\end{equation}
with ${\hat M}\equiv {\hat\Omega}(0_+) + {\hat\Omega}(0_-) - {\hat \Omega}(\beta) -
{\hat\Omega}(-\beta)$ and
\begin{equation}\label{eq:newdefom} {\hat \Omega}(\tau)\;\equiv\;
  \left( 
    \begin{array}{cc}
      \frac{1}{2} {\hat h}^{-1} 
      & \frac{i}{2} {\rm sgn}(\tau) \\
      -\frac{i}{2} {\rm sgn}(\tau) 
      & \frac{1}{2} {\hat h}
    \end{array}
  \right) \; e^{-{\hat h} |\tau|}  \;.
\end{equation}
Then,
\begin{equation}\label{eq:newm} 
{\hat M}\;\equiv\;
  \left( 
    \begin{array}{cc}
      {\hat h}^{-1} 
      & 0 \\
      0 
      & {\hat h}
    \end{array}
  \right) \;({\hat n}_B + 1)^{-1}\;,
\end{equation}
where
\begin{equation} 
{\hat n}_B \;\equiv\; \frac{1}{e^{\beta {\hat h}} - 1} \;.
\end{equation}
Coming back to the expression for ${\mathcal Z}_0(\beta)$, we see that:
\begin{eqnarray}\label{eq:zbose0} 
{\mathcal Z}_0(\beta) &=&
  \int {\mathcal D}\xi \,\exp \Big\{ 
  - \frac{1}{2} \int d^dx \int d^dy \big[ \xi_1 ({\mathbf x}) \;
  \langle {\mathbf x} |{\hat h}^{-1} \, ({\hat n}_B+1)^{-1} | {\mathbf y}
\rangle \;
  \xi_1
  ({\mathbf y}) \nonumber\\
  &+& \xi_2({\mathbf x}) \;\langle {\mathbf x} |{\hat h} \, ({\hat
n}_B+1)^{-1} |
  {\mathbf y} \rangle \; 
  \xi_2 ({\mathbf y}) \big] \Big\} \;.
\end{eqnarray}
By a simple field redefinition, we see that:
\begin{equation} 
{\mathcal Z}_0(\beta) \;=\; \det \big( {\hat n}_B + 1\big)
\end{equation}
which can be evaluated in the basis of eigenstates of momentum to yield:
\begin{equation} 
{\mathcal Z}_0(\beta) \;=\; \prod_{\mathbf k} \big[
  n_B(E_{\mathbf k}) + 1 \big]
\end{equation}
where $E_{\mathbf k} \equiv \sqrt{{\mathbf k}^2 + m^2}$. The free-energy
density, $F_0(\beta)$, is of course:
\begin{equation}
  F_0(\beta)\;=\; \frac{1}{\beta} \, \int \frac{d^dk}{(2\pi)^d} \, \ln
\big( 1
  \,-\, e^{-\beta E_{\mathbf k}}\big) \;.
\end{equation}

In the classical, high-temperature limit, the path integral for the
partition function becomes:
\begin{equation} 
{\mathcal Z}_0(\beta) \;\simeq\; \int {\mathcal D}\xi \,
e^{-\beta \; H(\xi)} \;,
\end{equation}
where:
\begin{equation}
  H(\xi) \;=\;\frac{1}{2} \int d^dx \big[ \xi_1^2({\mathbf x}) \,+\,
  |{\mathbf\nabla}\xi_2({\mathbf x})|^2 \,+\, m^2 \, \xi_2^2({\mathbf x}) 
  \Big] \;.
\end{equation}
This is, again, the usual classical expression for the partition function,
with the Lagrange multipliers playing the role of phase space variables,
and the integration measure being the corresponding Liouville measure. 
Besides, it is clear that the representation (\ref{eq:zbose0}) always
involves static fields, unlike in the Matsubara formalism. The price to pay
for this `dimensional reduction' is that the resulting `action' (the
exponent of the functional to be integrated) is spatially non local. It
becomes local only in the high-temperature limit.

\subsection{Quadratic approximation}\label{ssec:quad}
Let us study here a simple yet illuminating example where this approach
allows one to use non-perturbative $T=0$ information about a system as
input for the finite temperature partition function in a quite simple way. So we
assume that we know the exact $2$-point function ${\mathcal W}^{(2)}_{ab}$
at zero temperature. To be precise, since the Hamiltonian is
quadratic in the canonical momentum, we only need to know 
${\mathcal W}^{(2)}_{11}$, since it is possible to show that when one of
the fields is replaced by the canonical momentum, the result is multiplied
by the corresponding frequency (in Fourier space). Indeed, this can be
shown, for example, by performing the exact integral over the canonical
momentum.

Then the effective Hamiltonian corresponding to this term has the form:
\begin{equation}
	H_{eff}(\xi) \,=\, \frac{1}{2} \int_{{\mathbf x}, {\mathbf y}}
	\Big[ \xi_1({\mathbf x})
	C_1({\mathbf x}- {\mathbf y}) \xi_1({\mathbf y})
        \,+\, \xi_2({\mathbf x})
	C_2({\mathbf x}- {\mathbf y}) \xi_2({\mathbf y})\Big]
\end{equation}	
where the Fourier transforms of the coefficients $C_1$ and $C_2$ are
\begin{eqnarray}
	{\tilde C}_1 ({\mathbf k}) &=& \frac{1}{\beta}\, \int \frac{dk_0}{\pi} \;
\big( 1 - e^{ - i \beta k_0} \big)\, \tilde{\mathcal W}_{11}(k_0,{\mathbf
k}) \nonumber\\
	{\tilde C}_2 ({\mathbf k}) &=& \frac{1}{\beta}\, \int \frac{dk_0}{\pi} \;
\big( 1 - e^{ - i \beta k_0} \big)\, k_0^2 \,  \tilde{\mathcal
W}_{11}(k_0,{\mathbf k}) \;.
\end{eqnarray}
Note that these two coefficients that determine the contribution of the
quadratic term to the partition function, could be rather involved
functions of the functions of a coupling constant, since we are not
perturbing but just assuming that we only consider the information
contained in the full propagator. 

Assuming now that the exact propagator is such that only one particle, with
energy $E({\mathbf k})$, exists, one knows that there can only be poles
associated to them in $\widetilde{\mathcal W}_{11}$. Moreover, we assume that,
as usual, renormalization conditions have been imposed such that the
residue is $1$. Then we obtain:
\begin{eqnarray}
	{\tilde C}_1 ({\mathbf k}) &=& \frac{1}{\beta E({\mathbf k})} \,
	\big( 1 - e^{ - \beta E({\mathbf k})} \big) \nonumber\\
	{\tilde C}_2 ({\mathbf k}) &=& \frac{E({\mathbf k})}{\beta} \,
	\big( 1 - e^{ - \beta E({\mathbf k})} \big)\;, 
\end{eqnarray}
and the partition function becomes:
\begin{equation}
	{\mathcal Z}_s(\beta)\,=\, \prod_{\mathbf k} \frac{1}{1 - e^{-\beta
	E({\mathbf k})}} \;, 
\end{equation}
which is of course the one of an ideal Bose gas. But the important
difference is that one is obtaining it by putting the information
contained in the knowledge of the non-perturbative spectrum. It could be
the case, for example, of a model where the mass is generated by a
non-perturbative mechanism. 
Or one could have a theory with more than one pole, corresponding for
example to different bound states. These non-trivial poles, obtained in the
$T=0$ theory  are then directly taken into account by this contribution, as
the first terms in an expansion in powers of the auxiliary fields. Of
course, one has to include, in general, also non-quadratic terms. The
quadratic approximation could be justified, for example, within the context
of a large-$N$ approximation.

\subsection{Perturbation theory}\label{ssec:pert}
We study here the alternative representation for the partition function in an
interacting theory. For the sake of clarity, we present this topic within
the context of a concrete example: the real scalar field in $3+1$ dimensions,
with a self-interaction of the quartic type.  The action is then
\begin{equation}
 {\mathcal S} \;=\; {\mathcal S}_0 \,+\, {\mathcal S}_I \;\;,\;\;\;\;
{\mathcal S}_I\,=\, \frac{\lambda}{4!} \int d^4 x  \, \varphi^4(x) \;.
\end{equation}

As in the standard formulation, we want to calculate ${\mathcal Z}(\beta)$
in a power series of the coupling constant, $\lambda$. It should be clear
that the proper way to do that here is to calculate ${\mathcal W}[J]$ to the
desired order, and from there to obtain the corresponding effective
Hamiltonian. The resulting expression for the partition function as a
functional integral over the auxiliary fields can then also be expanded
(up to the same order ${\mathcal W}[J]$ was calculated). 
In the course of such a program one has to face the issue of the
UV infinites that pop up in the calculation of loop diagrams for the
(zero-temperature) object ${\mathcal W}[J]$. We
shall assume that all those infinities are renormalized in the usual way at
zero temperature. In particular, our $T=0$ vertices are normal ordered, so
that the zero-temperature tadpoles are,
to begin with, absent. The finite-temperature tadpoles will, nevertheless,
appear in the alternative description, as we shall see. 
Namely, the functional integral in terms of the auxiliary fields is UV finite.
In this way, the thermal corrections are completely disentangled from the issue of renormalization.  

We now study the problem of calculating the effective Hamiltonian in terms of
the perturbative expansion for ${\mathcal W}[J]$. As in the case of one
degree of freedom, the effective Hamiltonian is related to ${\mathcal
W}[J]$ by:
\begin{equation}
 H_{eff}(\xi) \;=\; -\frac{1}{\beta} \;{\mathcal
W}[i j(x)]
\end{equation}
where ${\mathcal W}[J]$ is the generating functional of connected
correlation functions $\Phi_a$, at $T=0$. Moreover, we shall assume that, in
the previous expression, ${\mathcal W}[0]\equiv 0$, since any vacuum
contributions would be cancelled by the normalization factor ${\mathcal N}$.

The perturbative expansion of ${\mathcal W}$ will be denoted by 
${\mathcal W}\,=\, {\mathcal W}^{(0)}+{\mathcal W}^{(I)}$, with
\begin{equation}
 {\mathcal W}_I[J] \;=\; {\mathcal W}^{(1)}[J]\,+\,{\mathcal W}^{(2)}[J]
\,+\, \ldots 
\end{equation}
where the index denotes the order in $\lambda$ of the corresponding term. 
This yields the corresponding expansion for the effective Hamiltonian,
$H_{eff}\,=\, H_{eff}^{(0)} + H_{eff}^{(I)}$, and one can then
find  corrections to the partition function, or the free energy 
$F= -\frac{1}{\beta} \ln {\mathcal Z}$, by evaluating the corresponding Gaussian averages. 
Indeed,
\begin{equation}
	F(\beta) \,= \, F^{(0)}(\beta)+ F^{(I)}(\beta)
\end{equation}
where
\begin{equation}
	F^{(I)}(\beta)\,=\,-\frac{1}{\beta} \ln \langle e^{-\beta
	H_{eff}^{(I)}(\xi)} \rangle
\end{equation}
where the average symbol is defined by the quadratic weight:
\begin{equation}
	\langle \ldots \rangle \;\equiv\; \frac{ \int {\mathcal D}\xi
	\ldots e^{-\beta H_{eff}^{(0)}(\xi)}}{{\mathcal Z}_0(\beta)} \;.
\end{equation}	

To fix ideas, we do that first for the simplest non-trivial order, i.e.,
$\lambda$. Then one has the first-order contribution to $F$:
\begin{equation}
	F^{(1)}(\beta)\,=\, \langle H_{eff}^{(1)}(\xi)\rangle \;,
\end{equation}
where the Gaussian average requires the knowledge of the elementary
averages which involve two auxiliary fields. The only non-trivial ones are:
\begin{eqnarray}
\langle \xi_1 ({\mathbf x}) \xi_1 ({\mathbf y})\rangle &=& 
\int \frac{d^3 k}{(2\pi)^3} e^{i {\mathbf k} \cdot ({\mathbf x} - {\mathbf
y})}
\frac{\omega({\mathbf k})}{ 1 - e^{ - \beta  \omega({\mathbf
k})}}\nonumber\\
\langle \xi_2 ({\mathbf x}) \xi_2 ({\mathbf y})\rangle &=& \int \frac{d^3
k}{(2\pi)^3} e^{i {\mathbf k} \cdot ({\mathbf x} - {\mathbf y})} 
\frac{1}{\omega({\mathbf k})( 1 - e^{ - \beta  \omega({\mathbf k})})}\;.
\end{eqnarray}

On the other hand, a standard $T=0$ calculation shows that the first-order
term in the expansion of ${\mathcal W}$ is: 
\begin{equation}
 {\mathcal W}^{(1)}[J] \;=\; - \frac{\lambda}{4!} \, \int d^4x \Big[ \int
d^4y \sum_a G^{(0)}_{1a}(x-y)  J_a(y) \Big]^4 \;,
\end{equation}
where $G^{(0)}_{ab}(x-y)$ is the free $T=0$ propagator for the real scalar
field and its canonical momentum, namely
\begin{equation}
G^{(0)}_{ab}(x-y)\,=\, \langle \Phi_a(x) \, \Phi_b(y) \rangle \;,
\end{equation}
which may be conveniently represented in terms of its Fourier
transforms ${\tilde G}^{(0)}_{ab}(k)$, in a matrix representation where ($1$
corresponds to $\varphi$ and $2$ to $\pi$):
\begin{equation}
{\tilde G}^{(0)}(k) \,=\,  
\frac{1}{k^2 + m^2} 
\left( 
 \begin{array}{cc}
  1 & i k_0 \\
 - i k_0 & k_0^2 
 \end{array}
\right) \;.
\end{equation}
Using now the rule that maps terms in the expansion for ${\mathcal W}$ into
like ones for $H_{eff}$, we see that:
\begin{eqnarray}\label{eq:heff1}
 H^{(1)}_{eff}(\xi) &=&  \frac{\lambda}{4! \beta} \, 
\int d^4x \Big[ \int d^4y \sum_a G^{(0)}_{1a}(x-y)  \nonumber\\ 
&\times & \xi_a({\mathbf y}) ( \delta(y_0 - \beta) - \delta(y_0) ) \Big]^4
\;.
\end{eqnarray}
Then, using tildes to denote the Fourier transforms of the auxiliary fields,
we see that each one of the factors that appear integrated over
$x$ above, can be put in the following form:
$$
\int d^4y \, \sum_a   G^{(0)}_{1a}(x-y) 
\xi_a({\mathbf y}) \big[ \delta(y_0 - \beta) - \delta(y_0) \big]
$$
\begin{eqnarray}
&=& \int \frac{d^3k}{(2\pi)^3} \, \frac{1}{2 \omega({\mathbf k})} \, 
e^{i {\mathbf k} \cdot {\mathbf x}}  \big[ e^{-\omega({\mathbf k})|x_0 -
\beta|} 
- e^{-\omega({\mathbf k})|x_0|} \big] {\tilde\xi}_1({\mathbf k}) 
\nonumber\\
&+& \int \frac{d^3k}{(2\pi)^3} \, e^{i {\mathbf k} \cdot {\mathbf x}} 
\frac{1}{2} \big[ \sigma(x_0) e^{-\omega({\mathbf k})|x_0|}
- \sigma(x_0-\beta)e^{-\omega({\mathbf k})|x_0 -\beta|} \big]
{\tilde\xi}_2({\mathbf k})\;,
\end{eqnarray}
where $\sigma(x)$ denotes the sign of $x$.

Introducing the expression above into (\ref{eq:heff1}) we find, in a natural
extension of the notation used for the case of one degree of freedom:
\begin{eqnarray}
H^{(1)}_{eff}(\xi) &=& \frac{1}{4!} \, 	\int   
\Big[\prod_{i=1}^4 \frac{d^3k_i}{(2\pi)^3}\Big]\,(2\pi)^3
\delta^{(3)}\big(\sum_{i=1}^4 {\mathbf k}_i\big)
\, 
{\mathcal H}^{(4,1)}_{a_1 \ldots a_4}({\mathbf k}_1, \ldots, {\mathbf k}_4)
\nonumber\\
&\times& {\tilde \xi}_{a_1}({\mathbf k}_1) \ldots {\tilde \xi}_{a_4}({\mathbf
k}_4)
\end{eqnarray}
where  ${\mathcal H}^{(4,1)}$ is the kernel for a quartic term; the
$`1'$ has been written to pinpoint the fact that it has been calculated to
the first order. The explicit form of the kernel elements depends of course
on the indices considered. Modulo permutations, the only inequivalent  
possibilities are summarized in the following results:
\begin{equation}
 {\mathcal H}^{(4,1)}_{1 1 1 1}({\mathbf k}_1, {\mathbf k}_2,
{\mathbf k}_3, {\mathbf k}_4) \,=\,\lambda \, \int 
\Big( \prod_{i=1}^4\frac{d \nu_i}{2\pi}\Big)  \, 
2\pi \, 
\delta(\sum_{j=1}^4 \nu_j) \, \prod_{l=1}^4 
\Big[\frac{e^{-i \beta \nu_l} - 1}{ \nu_l^2 + \omega^2({\mathbf k}_l)} \Big]
\;,
\end{equation}
\begin{eqnarray}
 {\mathcal H}^{(4,1)}_{1 1 1 2}({\mathbf k}_1, {\mathbf k}_2,
{\mathbf k}_3, {\mathbf k}_4) &=& \lambda\, \int 
\Big( \prod_{i=1}^4 \frac{d \nu_i}{2\pi}\Big)  \, 
2\pi \, \delta(\sum_{j=1}^4 \nu_j) \nonumber\\ 
&\times&
\prod_{l=1}^3 \Big[\frac{e^{-i \beta \nu_l} - 1}{ \nu_l^2 +
\omega^2({\mathbf k}_l)}\Big] 
\;\; \frac{i \nu_4 \big(e^{-i \beta \nu_4} - 1\big)}{ \nu_4^2 +
\omega^2({\mathbf k}_4)}
\end{eqnarray}
\begin{eqnarray}
 {\mathcal H}^{(4,1)}_{1 1 2 2}({\mathbf k}_1, {\mathbf k}_2,
{\mathbf k}_3, {\mathbf k}_4) &=& \lambda \, \int 
\Big( \prod_{i=1}^4 \frac{d \nu_i}{2\pi}\Big)  \, 
2\pi \, \delta(\sum_{j=1}^4 \nu_j) \nonumber\\ 
&\times&
\prod_{l=1}^2 \Big[\frac{e^{-i \beta \nu_l} - 1}{ \nu_l^2 +
\omega^2({\mathbf k}_l)}\Big] 
\;\; 
\prod_{r=3}^4 \Big[\frac{i \nu_r \big(e^{-i \beta \nu_r} - 1\big)}{ \nu_r^2 +
\omega^2({\mathbf k}_r)}\Big] \;,
\end{eqnarray}
\begin{eqnarray}
 {\mathcal H}^{(4,1)}_{1 2 2 2}({\mathbf k}_1, {\mathbf k}_2,
{\mathbf k}_3, {\mathbf k}_4) &=& \lambda \, \int 
\Big( \prod_{i=1}^4 \frac{d \nu_i}{2\pi}\Big)  \, 
2\pi \, \delta(\sum_{j=1}^4 \nu_j) \nonumber\\ 
&\times&
\frac{e^{-i \beta \nu_1} - 1}{ \nu_1^2 + \omega^2({\mathbf k}_1)} 
\;\; 
\prod_{r=2}^4 \Big[\frac{i \nu_r \big(e^{-i \beta \nu_r} - 1\big)}{ \nu_r^2 +
\omega^2({\mathbf k}_r)}\Big] \;,
\end{eqnarray}
and
\begin{equation}
 {\mathcal H}^{(4,1)}_{2 2 2 2}({\mathbf k}_1, {\mathbf k}_2,
{\mathbf k}_3, {\mathbf k}_4) \,=\, \lambda \, \int \Big( \prod_{i=1}^4 
\frac{d\nu_i}{2\pi}\Big) \, 
2\pi \, 
\delta(\sum_{j=1}^4 \nu_j) \,\prod_{l=1}^4 
\Big[\frac{i \nu_l \big(e^{-i \beta \nu_l} - 1\big)}{ \nu_l^2 +
\omega^2({\mathbf k}_l)}\Big] \;.
\end{equation}
Upon application of Wick's theorem for the calculation of the average of
the first order effective Hamiltonian, we see that only the terms with an
even number of legs for each field yields a non-vanishing contraction;
namely, only the terms ${\mathcal H}^{(4,1)}_{1 1 1 1}$, ${\mathcal
H}^{(4,1)}_{1 1 2 2}$, ${\mathcal H}^{(4,1)}_{2 2 2 2}$ enter into the
calculation. Besides, the first and third of these carry a factor of $3$
because of the number of unequivalent contractions, while for the second
one there is a $6$ due to the different permutations of the (different)
indices. 

Using the explicit form of the contractions, and integrating over the
frequencies, one sees, after a somewhat lengthy, but nevertheless
straightforward calculation, that the proper result is obtained. Namely,
\begin{equation}
	F^{(1)} \,=\, \frac{\lambda}{8 \beta} \, V \, 
	\Big(\int \frac{d^3k}{(2\pi)^3} \frac{1}{\omega({\mathbf k}) }
	n_B(\omega({\mathbf k}) )\Big)^{2} \;, 
\end{equation}
where $V$ is the spatial volume of the system.
\subsection{Generating functional}
Including a source in equation (\ref{eq:zesc1}) it is straightforward to
obtain the generating functional for the scalar field in $d+1$ dimensions
in a analogous form than for $0+1$ dimensional theory.  
Working in Fourier space for the spatial coordinates, we 
see that the entire $0+1$ dimensional procedure applies. 
Then $G_{ab}(x,y)$, the correlation function for the scalar field and its
canonical momentum becomes:
\begin{equation}\label{eq:ecuaProp} 
	G_{ab}(x,y)\,=\, G_{ab}^{(0)}(x-y)-U^{(\beta)}_{ab}(x-y)\;,
\end{equation}
where $G_{ab}^{(0)}(x-y)$ is the zero temperature correlation function, 
whereas $U_{ab}^{(\beta)}(x-y)$ is a temperature dependent function. 

The explicit form of these functions is:
\begin{equation}\label{eq:ecuaGzero} 
G_{ab}^{(0)}(x-y)\,=\, \int \frac{d^d \mathbf{k}}{(2 \pi)^d} \,e^{i
\mathbf{k}\,(\mathbf{x}-\mathbf{y})}\,\,
G_{ab}^{(0)}(\mathbf{k},\tau)
\end{equation}
where:
\begin{equation}
G_{ab}^{(0)}(\mathbf{p},\tau) \,=\,
\frac{e^{-\omega_{\mathbf k} \tau}}{2 \omega_{\mathbf k}}
\left( 
 \begin{array}{cc}
  1 & i \omega_{\mathbf k} \\
 - i \omega_{\mathbf k} & \omega^2_{\mathbf k} 
 \end{array}
\right) \;,
\end{equation}
and:
\begin{equation}
	U^{(\beta)}({\mathbf k},\tau) \,=\,- \frac{n_B(\omega_{\mathbf k})}{2}
\left( 
 \begin{array}{cc}
	 \frac{e^{\omega_{\mathbf k}\tau}+e^{-\omega_{\mathbf k}\tau}}{\omega_{\mathbf k}}
& -i \big(e^{\omega_{\mathbf k}\tau}-e^{- \omega_{\mathbf k}\tau}\big)\\ 
 - i \big(e^{ \omega_{\mathbf k} \tau}-e^{- \omega_{\mathbf k}\tau}
\big) &
\omega_{\mathbf k} \big(e^{\omega_{\mathbf k} \tau} + 
e^{-\omega_{\mathbf k}\tau} \big)
 \end{array}
\right) \;,
\end{equation}
which are exactly equal to the $0+1$ dimensional case equations
on which we have to replace $w$ by $\omega_{\mathbf k}=\sqrt{\mathbf{k}^2 +
m^2}$. 

\section{Dirac field}\label{sec:dirac}
The final example we consider is a massive Dirac field in $d+1$ spacetime
dimensions. The procedure will be essentially the same as for the real
scalar field, once the relevant kinematical differences are taken into
account. The action $S_0^f$ for the free case is given by $S_0^f =\int
d^{d+1}x \bar{\psi}(\not\!\partial + m)\psi$ where \mbox{$\not\!\partial=
  \gamma_{\mu}\partial_{\mu}$}, $\gamma_{\mu}^\dagger=\gamma_{\mu}$ and
$\{\gamma_{\mu},\gamma_{\nu}\}=2 \delta_{\mu\nu}$.

We then impose antiperiodic conditions for both fields:
\begin{equation}
  \psi\left(\beta,{\mathbf x}\right) \,=\, -{\psi} 
  \left(0,{\mathbf x}\right)\;,\;\;
  \bar{\psi}\left(\beta,{\mathbf x}\right) \,=\, -\bar{\psi} 
  \left(0,{\mathbf x}\right)
\end{equation}
as constraints on the Grassmann fields. Those conditions lead to the
introduction of the two $\delta-$functions:
\begin{eqnarray} {\mathcal Z}_0^f(\beta) &=& \int \, {\mathcal
D}\psi{\mathcal
    D}\bar{\psi}\, \delta \big(\psi(\beta,{\mathbf x})+\psi(0,{\mathbf
    x})\big)\; \delta \big(\bar{\psi}(\beta,{\mathbf x})
  +\bar{\psi}(0,{\mathbf x})\big) \nonumber\\
  &\times& \exp \Big[-S_0^f(\bar{\psi},\psi)\Big] \;.
\end{eqnarray}
Since the Dirac action is of the first-order, the introduction of two
constraints, and two Lagrange multipliers, appears in an even more natural
way than for the previous case.  Those auxiliary fields, denoted by
$\chi(\mathbf
x)$ and $\bar\chi(\mathbf x)$ must be time-independent Grassmann spinors. 
The resulting expression for ${\mathcal Z}_0^f(\beta)$ is then
\begin{equation} 
	{\mathcal Z}_0^f(\beta) \;=\; {\mathcal N}^{-1}\, \int \,
{\mathcal
    D}\chi{\mathcal D}\bar{\chi}{\mathcal D}\psi{\mathcal D}\bar{\psi}\,
  e^{-S_0^f(\bar{\psi},\psi)+ i \, \int d^{d+1}x \,(\bar{\eta}\psi +
    \bar{\psi}\eta )},
\end{equation}
where $\eta$ and $\bar{\eta}$ are (Grassmann) sources depending on $\chi$
and $\bar{\chi}$ through the relations:
\begin{equation}
  \eta(x)\,=\,\chi({\mathbf x})\big[\delta(\tau -\beta)
  +\delta(\tau)\big]\;,\;\;\;\bar{\eta}(x)\,=\,\bar{\chi} ({\mathbf
    x})\big[\delta(\tau -\beta)+\delta(\tau)\big]\;.
\end{equation}

Integrating out $\psi, \bar{\psi}$, we arrive to:
\begin{equation} 
{\mathcal Z}_0^f(\beta)\;=\; \int \, {\mathcal
    D}\chi{\mathcal D}\bar{\chi}\, \exp \Big[-\beta
  H_{eff}\big(\bar{\chi},\chi \big)\Big]
\end{equation}
where
\begin{equation}
  H_{eff}\big(\bar{\chi},\chi \big)= \int d^{d}x \,\int d^{d}y \,
  \bar{\chi}({\mathbf x})H^{(2)}\big({\mathbf x},{\mathbf y}\big
)\chi({\mathbf y})
\end{equation}
with:
$$
  H^{(2)}\big({\mathbf x},{\mathbf y}\big ) \;=\; \langle \mathbf x,0|
  (\not\!\partial + m)^{-1}|\mathbf y,0 \rangle + \langle\mathbf
x,\beta|(\not\!\partial 
  + m)^{-1}|\mathbf y,\beta \rangle 
$$
$$
+\, \langle \mathbf x,0|(\not\!\partial + m)^{-1}|\mathbf y,\beta \rangle
+ \langle \mathbf x,\beta|(\not\!\partial + m)^{-1}|\mathbf y,0
\rangle
$$
\begin{equation}
=\, \frac{1}{\beta}\,\Big[ 2\,S_f\big(0,\mathbf x - \mathbf y\big) 
  + S_f\big(\beta,\mathbf x - \mathbf y\big) 
  + S_f\big(-\beta,\mathbf x - \mathbf y\big)\Big].
\end{equation}
On the last line, $S_f$, denotes the Dirac propagator.  A quite
straightforward calculation shows that
\begin{equation}
  H\big({\mathbf x},{\mathbf y}\big )=\frac{1}{\beta}\,
  \langle\mathbf x|\hat{u}(1 - \hat{n}_F)^{-1}|\mathbf y\rangle
\end{equation}
where $\hat{n}_F \equiv \Big( 1+ e^{\beta \hat{n}}\Big)^{-1}$ is the
Fermi-Dirac distribution function, written in terms of
\mbox{$\hat{h}$}, the energy operator (defined identically to its real
scalar field counterpart);  $\hat{u}$ is a unitary operator, defined as
\begin{equation}
  \hat{u} \;\equiv\; \frac{\hat{h}_D}{\hat{h}}\, , \;\; \hat{h}_D \equiv
  {\mathbf\gamma}\,\cdot\,{\mathbf \nabla} + m \;.
\end{equation}  

Then we verify that:
\begin{equation} 
{\mathcal Z}_0^f(\beta) \;=\;\det\hat{u}\;
{\det}^{-1}\big[
  (1-\hat{n}_F)\,\mathbf I\big] \;,
\end{equation}
($\mathbf I\,\equiv$ identity matrix in the representation of
Dirac's algebra)
\begin{equation} 
{\mathcal Z}_0^f(\beta) \;=\;\left\{\prod_{\vec{p}} \Big[
1 + e^{-\beta E(\vec{p})}\Big] \right\}^{r_d}
\end{equation}
with $E({\mathbf p})=\sqrt{{\mathbf p}^2 + m^2}$ and $r_d\,\equiv$
dimension of the representation (we have used the fact that $\det\hat{u} =
1$).

Again, the procedure has produced the right result for the partition
function, with a normal-ordered Hamiltonian. On the other hand, for a Dirac
field in a static external background corresponding to a minimally coupled
Abelian gauge field the $A_{0}=0$ gauge, we have
\begin{equation}\label{eq:defsDA}
  S^f (\bar{\psi},\psi, A)\;=\;\int d^{d+1}x \,
  \Big[ \bar{\psi}(x)\big(\not\!\partial + i\,e\, 
  {\mathbf\gamma}\cdot {\mathbf A} ({\mathbf x}) + m \big)\psi(x) \Big] \;.
\end{equation}
The assumed $\tau-$ independence allows us to carry on the derivation
described for the free case, with minor changes, arriving to the
expression:
\begin{eqnarray} 
{\mathcal Z}^f(\beta) \;&=&\;\det\hat{u}({\mathbf A})\,
  {\det}^{-1}\big(\hat{n}_{F}({\mathbf A})\,\mathbf I\big)\nonumber\\
  &=& e^{i K({\mathbf A})} \det\Big[\big( 1 + e^{-\beta \hat{h}({\mathbf
A})
  }\big)\mathbf I\Big]
\end{eqnarray}
where
\begin{equation}
  \hat{h}(A)\equiv \sqrt{-{\mathbf D}^2+m^2}\;,\;\;
  {\mathbf D}\equiv {\mathbf\nabla}-i e {\mathbf A} \;,
\end{equation}
and:
\begin{equation}
  e^{i K({\mathbf A})}\;=\;\frac{\det\big( {\mathbf\gamma}\cdot {\mathbf D}
+ m \big)}{\det   \sqrt{-{\mathbf D}^{2} + m^{2}}}\;.
\end{equation}
Notice that the factor \mbox{$\det\Big[\big( 1 + e^{-\beta \hat{n}({\mathbf
   A})}\big) \mathbf I\Big]$} can be formally diagonalized in terms of
the energies $E_\lambda({\mathbf A})$ in the presence of the external
field. Thus we arrive to the expression:
\begin{equation} 
{\mathcal Z}^f(\beta) \;=\;e^{i K({\mathbf A})}\;\times \;
  \left\{\prod_\lambda \Big[ 1 + e^{-\beta E_\lambda({\mathbf A})}\Big]
  \right\}^{r_d} \;.
\end{equation}
The factor $e^{i K({\mathbf A})}$, on the other hand, is topological in
origin, as it depends on the phase, $K({\mathbf A})$, of the determinant of
$\hat{h}_D$. On the other hand, $\hat{h}_D$ may be regarded as a kinetic
operator in one fewer dimension.  For Dirac fermions, we know that the
phase of $\det\hat{h}_D$ can be non-trivial only when $d$ is odd, i.e.,
when $d+1$ is even. However, the $\gamma$-matrices appearing in
$\det\hat{h}_D$ form a reducible representation of the Dirac algebra in $d$
dimensions, with the matrix $\gamma_\tau$ relating every eigenvalue to its
complex conjugate. Thus, as a result, the phase $K({\mathbf A})$ vanishes.
Of course, a non-vanishing result may be obtained for other fermionic
systems, like Weyl fermions for \mbox{$d+1={\rm even}$}.
\section{Conclusions}\label{sec:concl}
 We have shown that, by introducing the periodicity conditions 
as constraints for the paths in the Euclidean path integral for the 
$T=0$ vacuum functional,
one can obtain a novel representation for the partition function. These
constraints should be applied on fields and canonical momenta, and when they
are represented by means of auxiliary fields, they lead to an alternative,
`dual' representation for the corresponding thermal observable.

Since both phase space variables should be constrained, one has to work in
a first-order formulation; this is automatically satisfied in the case of a
Dirac field, but it requires a little bit of care in the case of the real
scalar field.

The resulting representation for the partition function may be thought of
as a dimensionally reduced path integral over phase space, similar to the
one of a classical thermal field theory, with the auxiliary fields playing the
role of canonical variables, but with an effective Hamiltonian, $H_{eff}$, which 
reduces to the classical one in the corresponding (high-temperature) limit.  

We analyzed the main properties of this representation for the cases of the
real scalar and Dirac fields, two typical examples that have been chosen
for the sake of simplicity. It is not difficult to generalize the
representation to the case of systems containing  fermions interacting with
bosons. For example, assuming that $S({\bar\psi},\psi,\Phi)$ is the first
order action corresponding to a real scalar interacting with a Dirac field,
we define the $T=0$ generating functional \mbox{${\mathcal W}({\bar
\zeta},\zeta, J)$} by:
\begin{equation}
{\mathcal W}({\bar \zeta},\zeta, J) \;=\; {\mathcal N}\int \,  
{\mathcal D}\Phi {\mathcal D}\psi{\mathcal D}\bar{\psi}\, 
e^{-S(\bar{\psi},\psi,\Phi)+ i \,\int d^{d+1}x \,\big( \bar{\zeta}\psi + \bar{\psi}\zeta  
+ J_a \Phi_a \big)},
\end{equation}
where the sources are arbitrary. Then, $H_{eff}$ can be obtained from the
expression:
\begin{equation}
H_{eff}({\bar\chi},\chi,\xi) \;=\; -\frac{1}{\beta} \;{\mathcal W}({\bar
\eta},\eta, i j) \;, 
\end{equation}
where $\eta$, ${\bar \eta}$ and $J$  are (the already defined) functions of
the Lagrange multiplier fields ${\bar\chi}$, $\chi$ and $\xi$.

We have shown how the effective Hamiltonian can be constructed by assuming
the knowledge of the corresponding $T=0$ generating functional of connected
correlation functions. If this knowledge is perturbative, one recovers the
perturbative expansion for the thermal partition function. However, the
most important applications of this formalism are to be found in the case
of having non-perturbative information about the $T=0$ correlation
functions: here, it is quite straightforward to incorporate that knowledge
into the formalism, and to compute thermal corrections from it.

\section*{Acknowledgements}
C.D.F. and C.C.T. thank CONICET (PIP5531) and CAPES/SPU for financial
support.  A.P.C.M.  and I.R. thank CAPES/SPU, CNPq/MCT and FAPERJ
for partial financial support.

\end{document}